\documentclass[conference]{IEEEtran}
\IEEEoverridecommandlockouts
\usepackage{cite}
\usepackage{amsmath,amssymb,amsfonts}
\usepackage{graphicx}
\usepackage{booktabs}
\usepackage{float}
\usepackage{tabularx}
\usepackage{caption}
\usepackage{subcaption}
\usepackage{breqn}
\usepackage[export]{adjustbox}
\usepackage{hyperref}  	
\hypersetup{
    allcolors=black,
    colorlinks=true,
    linkcolor=black,
    urlcolor=blue,
}

\def\BibTeX{{\rm B\kern-.05em{\sc i\kern-.025em b}\kern-.08em
    T\kern-.1667em\lower.7ex\hbox{E}\kern-.125emX}}
    
\newcolumntype{C}{>{\centering\arraybackslash}X}
\newlength\myindent
\begin{document}

    \title{Gotta Catch 'em All: Aggregating CVSS Scores\\}
    
    \author{\IEEEauthorblockN{1\textsuperscript{st} Ángel Longueira-Romero, 2\textsuperscript{nd} Jose Luis Flores, 3\textsuperscript{rd} Rosa Iglesias}
    \IEEEauthorblockA{\textit{Industrial Cybersecurity} \\
    \textit{Ikerlan Technology Research Centre (BRTA)}\\
    Arrasate/Mondragón, Spain \\
    \{alongueira, jlflores, riglesias\}@ikerlan.es}
    \and
    \and
    \IEEEauthorblockN{4\textsuperscript{th} Iñaki Garitano}
    \IEEEauthorblockA{\textit{Dept. of Electronics and Computing} \\
    \textit{Mondragon Unibertsitatea}\\
    Arrasate/Mondragón, Spain \\
    igaritano@mondragon.edu}
    }
    
    \maketitle
    
    \begin{abstract}
    Security metrics are not standardized, but international proposals such as the Common Vulnerability Scoring System (CVSS) for quantifying the severity of known vulnerabilities are widely used. Many CVSS aggregation mechanisms have been proposed in the literature. Nevertheless, factors related to the context of the System Under Test (SUT) are not taken into account in the aggregation process; vulnerabilities that in theory affect the SUT, but are not exploitable in reality. We propose a CVSS aggregation algorithm that integrates information about the functionality disruption of the SUT, exploitation difficulty, existence of exploits, and the context where the SUT operates. The aggregation algorithm was applied to OpenPLC V3, showing that it is capable of filtering out vulnerabilities that cannot be exploited in the real conditions of deployment of the particular system. Finally, because of the nature of the proposed algorithm, the result can be interpreted in the same way as a normal CVSS.
\end{abstract}

\begin{IEEEkeywords}
CVSS, security metrics, aggregation, attack graphs, vulnerabilities.
\end{IEEEkeywords}
\section{Introduction}
    System security quantification is not an easy task~\cite{securityHard}. There exist both a lack of consensus and standardization around security metrics~\cite{whySecurityTesting_Herbert, whySecurityMetrics_Atzeni, weakHypothesis_Verendel, securityMeasuring_Stolfo, holyGrail_intro_1, criticalSecurityIndicators_Rudolph, Sentilles2018WhatDW}. For this reason, research efforts keep aiming to unify this field~\cite{securityMetrics:indin2020}.

    Among these efforts, the Common Vulnerability Scoring System (CVSS) is a widely extended standard for vulnerability quantification~\cite{CVSS1}. CVSS is a public framework that provides a standardized method for assigning quantitative values to security vulnerabilities according to their severity. A CVSS score is a decimal number in the range [0, 10]\footnote{The latest version at the time this paper was written is version 3.1.}~\cite{CVE2}.

    The CVSS is aimed to quantify the severity of vulnerabilities in individual and specific software items, however the majority of systems are actually a composition of simpler isolated items with different interdependencies. This situation highlights one of the biggest problems related to security quantification~\cite{Pendleton_OverallvsGranularity}, the difficulty to really measure the global security state of a composite system. To do so, it would be necessary to aggregate each individual CVSS value into a global one in a consistent and coherent way.

    The official CVSS documentation does not propose any kind of aggregation mechanism, and nowadays, there is no standardized method~\cite{CVSS_Broken}. In addition to this, previous research works do not usually integrate contextual or interdependency information about the vulnerabilities to update the CVSS. This means that aspects such as whether affected functionalities, the environment of deployment, or the existence of exploits are usually neglected.

    Context is a critical aspect to integrate in the aggregation process. This can be illustrated using a device implementing multiple functionalities as an example. To perform those functionalities, usually it will contain assets that implement those functionalities. But depending on the context where the device is deployed, some of its functionalities might not be needed. So the assets implementing unused functionalities would be disabled, and therefore, their vulnerabilities could not be exploited. It can also be the case that the asset implementing a functionality is simply inaccessible, so it could not also be exploited.

    This research proposes a novel aggregation algorithm for a set of CVSS values\footnote{The Python code implementing the aggregation algorithm is available at GitHub \url{https://github.com/aaalongueira/CVSS_Aggregation}.}. This approach is based on the Extended Dependency Graphs (EDGs) proposed by Longueira-Romero \textit{et al.}~\cite{Longueira-Romero_EDG_2022}. Because EDGs are capable of modeling dependencies, this algorithm can also be applied to computer networks. Our proposal is capable of selecting the most relevant CVSS to be aggregated, taking into account four different context-related properties of the System Under Test (SUT):
    \begin{enumerate}
        \item Functionality disruption.
        \item Exploitation difficulty.
        \item Existence of exploits, and their development state.
        \item Context of deployment.
    \end{enumerate}

    This approach increases the granularity of the CVSS base, environment and temporal metrics, where not every possible value in the scale $[0,10]$ is achievable, or the result of changing the value of a submetric has almost no effect on the final CVSS~\cite{CVSS_Broken, CVSS_update}. Moreover, our proposal is capable of detecting which branch in the EDG is contributing the most (more critical) to the final score.

    This paper is organized as follows: We review existing aggregation methods in Section~\ref{sec:relatedWork}. Our proposal is explained in Section~\ref{sec:proposal}, and tested in a use case in Section~\ref{sec:useCase}. Finally, Section~\ref{sec:conclusions} contains the conclusions and future work of this research.
\section{Related Work}
\label{sec:relatedWork}
    Nowadays, there is no widely-accepted method to aggregate CVSS values for software composition. All of them can be classified into one of the following categories~\cite{aggregatingMethodsClassification_1, aggregatingMethodsClassification_2}: 
    (1) Arithmetic Aggregation, 
    (2) Attack Graph-based Aggregation, and 
    (3) Bayesian Network-based Aggregation.

    \subsection{Arithmetic Aggregation}
        This method uses arithmetic operations to aggregate the values~\cite{naive_1, naive_2, naive_3, naive_4}. Common examples of this approach are taking the maximum of the CVSS values, their arithmetic mean, or a combination of them. For example, Heyman \textit{et al.}~\cite{naive_1}, proposed an algorithm to aggregate CVSS values in dependency graph that is based on taking the maximum value in each case, according to certain conditions.

        Although their simplicity makes them suitable for initial approximations, their results can be biased in two ways:
        \begin{enumerate}
            \item \textbf{Exploitable by quantity:} When a system poses several vulnerabilities that by their own are not critical and cannot be exploited, they can sum up to an aggregated value of a high impact vulnerability (overfitting). This can happen when multiple simple mechanisms are combined as the aggregation algorithm.

            \item \textbf{Exploitable by criticality:} When there exist a critical vulnerability, the whole system will be usually classified as critical. Nevertheless, that vulnerability might not be exploitable, nor being affecting the functionality of the system. This is specially common when using the maximum as the aggregation algorithm.
        \end{enumerate}

    \subsection{Attack Graph-based Aggregation}
        This approach models the relationships between vulnerabilities using attack graphs, converting CVSS scores into probabilities~\cite{attackGraph_1, attackGraph_2, attackGraph_3, attackGraph_4, attackGraph_5, attackGraph_6}. In this way, both the CVSS value and the place of the vulnerability in the whole graph are taken into account.
        
        Cheng \textit{et al.} in~\cite{aggregatingMethodsClassification_1} proposed a graph-based aggregation method that uses the underlying metrics of CVSS, where the dependency relationships between vulnerabilities are usually visible. As the center of the aggregation algorithm, they use the product of the CVSS used as probabilities, also known as the join probability of both vulnerability.

        The main drawback with these approaches is that the relationship between individual vulnerabilities cannot be obtained straightforwardly from existing databases. This means that establishing a relation between two vulnerabilities implies that they can be chained during an attack, which is not always obvious. Moreover, factors such as exploitability of the vulnerabilities, or existing exploits are not taken into account.

    \subsection{Bayesian Network-based Aggregation}
        Going a step further, these methods integrate the conditional relationship between vulnerabilities, modeling them using Bayesian networks~\cite{bayesianNetwork_1, bayesianNetwork_2, bayesianNetwork_3}. Poolsappasit \textit{et al.}~\cite{bayesianNetwork_2} proposed a CVSS aggregation framework using Bayesian networks. They used the Bayesian probability factorization formula as the aggregation mechanism:
        
        \begin{equation*}
            p( x ) = \prod\limits_{i=0} p( x_v | x_{pa( v )} )
        \end{equation*}

        Bayesian network-based approaches have to deal with establishing the relationships between the vulnerabilities, but also with the calculation of conditional probabilities, that have to be usually estimated. As the previous ones, these techniques do not integrate information about how functionality if affected by existing vulnerabilities, or the possibility to actually exploit them.
\section{Proposed Approach for Metric Aggregation}
\label{sec:proposal}
    In this paper, we propose a CVSS aggregation algorithm inspired by the risk propagation formula~\cite{MAGERIT_III} described in MAGERIT~\cite{MAGERIT_I, Syalim_ComparisonRiskMethodologies_2009}. First, we describe the corrections factors involved in our proposal. Then, the aggregation formula is introduced. Finally, the algorithm and the interpretation of the results is explained in detail.
    
    \subsection{Correction Factors}
        The proposed aggregation algorithm integrates correction factors to adapt the formula described in MAGERIT. These correction factors apply individually for each CVSS, except for the average and summarized factors. Correction factors are summarized in Table~\ref{tab:correctionFactors}.

        \begin{table*}[!htb]
            \caption{Correction factors proposed for adapting the Bayesian sum proposed in MAGERIT.}
            \label{tab:correctionFactors}
            \centering
            \begin{tabularx}{\textwidth}{lXl}
                \hline
                CORRECTION FACTOR               & DESCRIPTION & AUTOMATED\\
                \hline
                Functionality factor ($\rho$)   & Binary value indicating whether a vulnerability affects or not the functionality of the SUT. & $\blacksquare$\\
            
                Deepness factor ($\beta$) & Value between $\left[0, 1\right]$ proportional to the position of the affected asset in the EDG of the SUT. & $\blacksquare$\\

                Context factor ($\gamma$)       & Binary value indicating vulnerability exploitability in the real and particular conditions of the SUT. & $\square$\\ 
            
                Exploit factor ($\mu$)          & Existence of a public exploit, proportional to its state of development: Not defined ($\mu = 0.5$), Theoretical ($\mu = 1.25$), Proof-Of-Concept ($\mu = 1.5$), Functional ($\mu = 1.75$), and Automated ($\mu = 2$).  & $\blacksquare$\\ 

                Summarized factor ($\lambda$)   & This factor summarizes the effect of all the above ones, $\lambda = \rho\beta\gamma\mu\sigma$. & $\blacksquare$\\

                Average factor ($\sigma$)       & Function that adjust the value of the sum to avoid its rapid evolution to 10. & $\blacksquare$\\
                \hline
            \end{tabularx}%
        \end{table*}
    
        \begin{enumerate}

            \item \textit{Functionality factor ($\rho$):}
                This correction factor represents whether any functionality of the systems is affected by its vulnerabilities. It is represented by a binary value, being $0$ when no functionality is affected, and $1$ when any of them is affected. For example, a cryptographic library with a vulnerability in SHA1. If the SUT does not make use of SHA1 in any way, the vulnerability would not be exploitable, and could be removed from the analysis ($\rho = 0$).
    
            \item \textit{Deepness Factor ($\beta$):}
                This factor represents the difficulty of chained exploitation of each vulnerability. It is represented by a value between $\left[0, 1\right]$ inversely proportional to the amount of assets to compromise in order to exploit vulnerability. Vulnerabilities close to the entry point will account more for the final aggregation, whereas those that are far away will account less. In this approach, linear interpolation is proposed to calculate the weight of each layer, because of its simplicity. Nevertheless, different interpolations could be used  according to the criticality of the system. Fig.~\ref{fig:deepnessFactor} shows the corresponding $\beta$ for a four-layer system.

                \begin{figure}[!htb]
                    \centering
                    \begin{subfigure}[t]{0.4\textwidth}
                        \centering
                        \includegraphics[width=0.2\textwidth,valign=c]{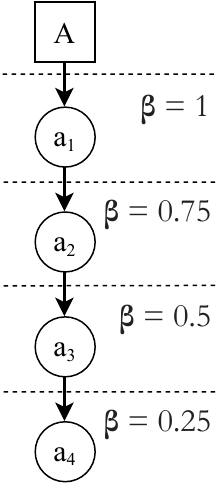}
                        \hfill
                        \includegraphics[width=0.6\textwidth,valign=c]{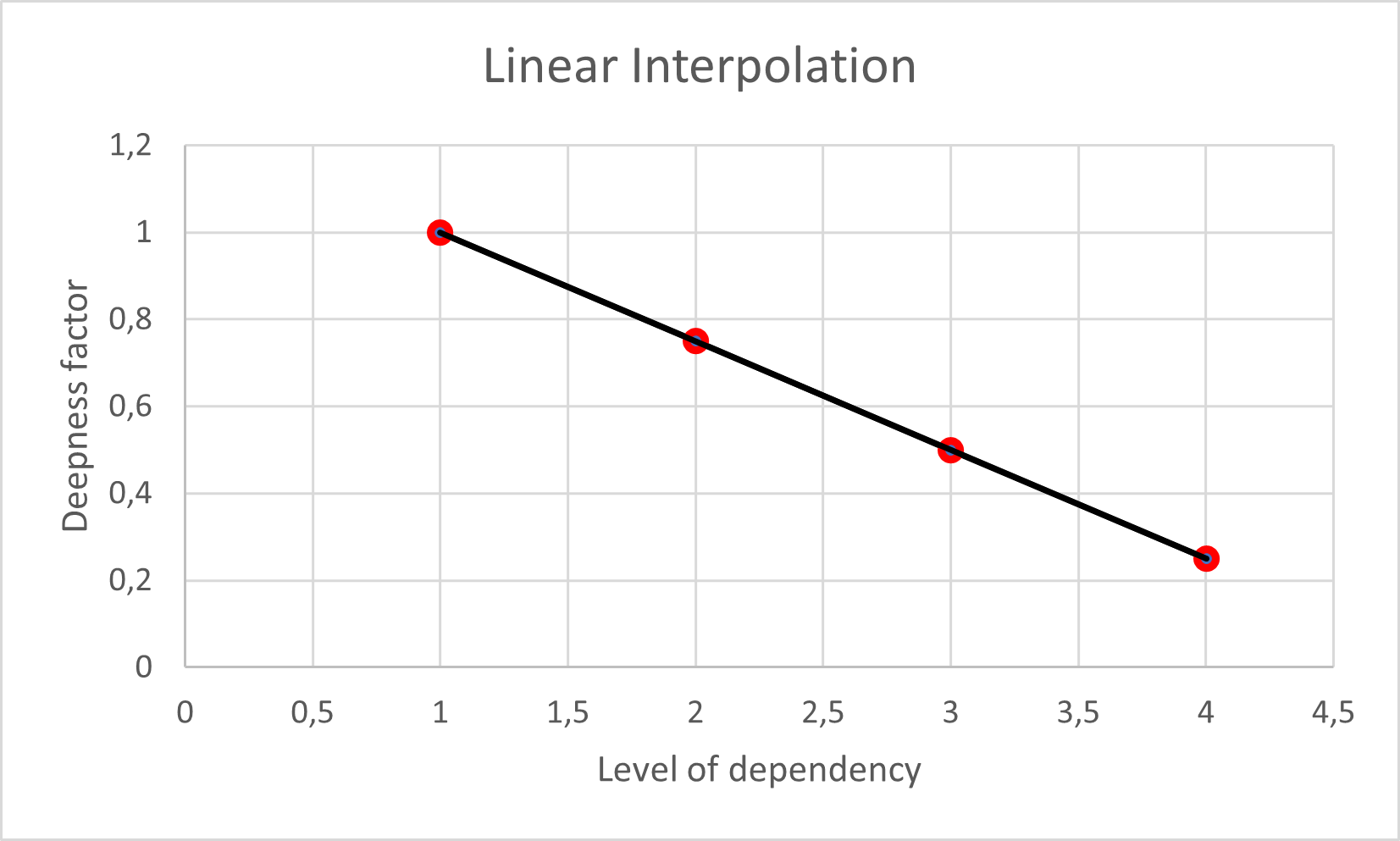}
                    \end{subfigure}
                    \caption{Calculation of the deepness factor for a four-layer of dependency example.}
                    \label{fig:deepnessFactor}
                \end{figure}

            \item \textit{Context factor ($\gamma$):}
                This factor considers whether the exploitation of a vulnerability is actually possible in the real scenario where the system is deployed. It is represented by a binary value, where $0$ indicated that it is not possible, and $1$, that it is possible. It is calculated comparing the attack vector of the CVSS with the real conditions where the device is deployed. For example, this can happen when a vulnerability with a high CVSS score needs physical access to be exploited, but in reality the device is physically isolated. To reflect this, the CVSS should be updated, lowering the resulting value~\cite{aggregatingMethodsClassification_1}. This factor aims to complement the existing submetrics in the temporal and the environment metrics of the CVSS. Both the temporal and the environmental scores lack of an "isolated" value for the attack vector.

            \item \textit{Exploit factor ($\mu$):}
                This factor accounts for the existence of a public exploit for a given vulnerability, being proportional to its state of development. The temporal score of the CVSS already implements this feature, but the CVSS values are not updated in practice~\cite{CVSS_update}. Moreover, taking into account the temporal score has almost no effect as opposed to using the raw initial base score. This means that a CVSS just considering the base score is higher than a CVSS considering an exploit code maturity of ``functional exploit exists''. To solve this issue, we introduce the following values for the exploit factor: 
                    Not defined ($\mu = 0.5$), 
                    Theoretical ($\mu = 1.25$), 
                    Proof-Of-Concept ($\mu = 1.5$), 
                    Functional ($\mu = 1.75$), and 
                    Automated ($\mu = 2$). 
                These values are equivalent to the scale defined in the CVSS Specification Document~\cite{CVSS1}.
    
            \item \textit{Summarized factor ($\lambda$):}
                The $\lambda$ factor accounts for the effect of all the factors above:
                \begin{equation}
                \label{eq:summarizedFactor}
                    \lambda = \rho\beta\gamma\mu
                \end{equation}
    
            \item \textit{Average factor ($\sigma$):}
                This factor defines the behavior of the aggregation function. It can be chosen as needed (\textit{e.g.}, the arithmetic or harmonic mean), but taking into account all the values to be added.
        \end{enumerate}
    
    \subsection{Aggregation Formula}
        The aggregation function is defined as:
        
        \begin{equation}
        \label{eq:addition}
            \Gamma(\overrightarrow{V}) = 10 - \frac{1}{\sigma} f(\overrightarrow{V})
        \end{equation}

        Where $\overrightarrow{V}$ is a vector $\left( cvss_0, cvss_1, \dots, cvss_n \right)$ with all the corrected CVSS values to be added, $cvss$, being $n$ the last value to be added. $f(\overrightarrow{V}) = a_n$ is defined as the following recursive function:
        
        \begin{dmath}
            a_n = 10\left[ 1 - \left( 1 - \frac{\lambda_{a_{n-1}}}{10} a_{n-1} \right)\cdot \left( 1 - \frac{\lambda_{cvss_n}}{10} cvss_n \right) \right]
        \end{dmath}

        Where the base case is defined as:
        
        \begin{equation}
            a_0 = \lambda_{cvss_0}cvss_{0}
        \end{equation}

    \subsection{Algorithm}
\begin{figure}[!htb]
          \begin{center}
          \includegraphics[width=0.25\textwidth]{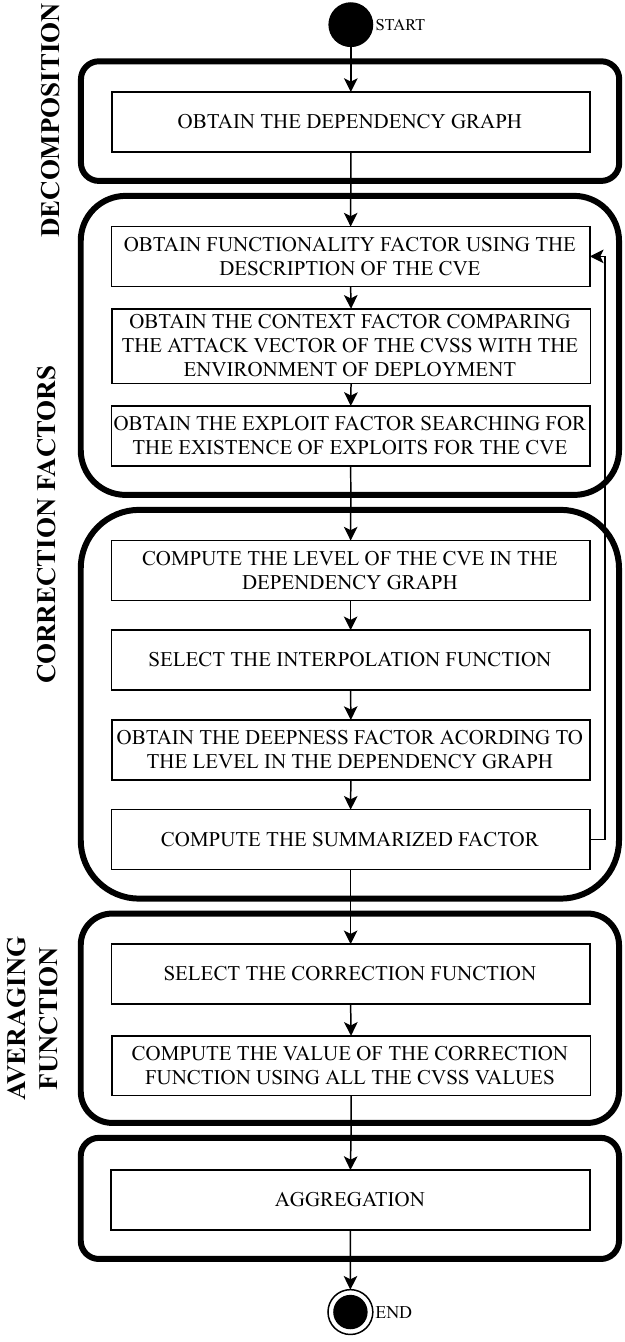}
          \caption{Flowchart showing the main steps of the aggregation algorithm for each CVSS.}
          \label{fig:aggregationAlgorithm}
          \end{center}
        \end{figure}

        The proposed aggregation algorithm is divided into the following steps (see Fig.~\ref{fig:aggregationAlgorithm}):
        \begin{enumerate}
            \item Calculation of the correction factors for each CVSS, 
            \item Calculation of the summarized factor for each CVSS, 
            \item Calculation of the corrected CVSS values,
            \item Calculation of the average correction function, and
            \item Aggregation.
        \end{enumerate}

        Notice that the dependency graph of the SUT, the vulnerabilities associated to each element of the dependency graph, and their CVSS value are needed.

        \subsubsection{Correction factors for each CVSS}
            The first step obtains the values of each correction factor for each CVSS:
        \begin{enumerate}
            \item \textbf{Functionality factor ($\rho$):} This factor is obtained using the description provided in the corresponding CVE of each CVSS. The description provides enough information to decide whether the functionality of the system is affected.

            \item \textbf{Context factor ($\gamma$):} This factor is obtained by comparing the value of the Attack Vector (AV) submetric of the CVSS, with the real environment of deployment of the SUT.

            \item \textbf{Exploit factor ($\mu$):} To obtain this factor, public databases have to be queried to find any potential exploit for each vulnerability.

            \item \textbf{Deepness factor ($\beta$):} For any given CVSS, its value is obtained according to the deepness in the exploit chain for the SUT~\cite{Longueira-Romero_EDG_2022}.
        \end{enumerate}

        \subsubsection{Summarized factor for each CVSS}
            The summarized factor, $\lambda$, is obtained by multiplying all the corrections factors obtained in the previous step, following Equation~\ref{eq:summarizedFactor}.

        \subsubsection{Corrected CVSS values}
            The corrected CVSS values are obtained by multiplying each CVSS by its corresponding summarized factor, $\lambda$. At this point, it is necessary to check for overflows, because the exploitation factor generated corrected CVSS values higher than 10. Values higher than 10 are set to 10 at this stage.

        \subsubsection{Correction function}
            At this point, it is necessary to choose an averaging function. Choosing one function over the other will cause the aggregation result to grow slower or faster toward 10 in each addition. In this case, and for the sake of clarity, we chose the arithmetic mean, but any other kind of mean (\textit{e.g., harmonic mean}) could be used according to each scenario.

        \subsubsection{Aggregation}
            Finally, the aggregated value is computed using Equation~\ref{eq:addition}.

    \subsection{Interpretation of the result}
        The advantage of this method is that the result can be interpreted in the same way that a normal CVSS would be interpreted. This is because of the correction factors in Equation~\ref{eq:addition}, that only let the algorithm return high values when vulnerabilities with high CVSS values are exploitable in reality ($\lambda$ is close to $1$). This mechanism ensures that multiple aggregated low CVSS values do not result in a critical score just because there are a large number of them.

\section{Use Case}
\label{sec:useCase}
    To test the potential of our proposal, we analyzed Version 3 of OpenPLC project, obtaining a CVSS aggregated value for its vulnerabilities using the proposed algorithm.

    OpenPLC is the first functional open source Programmable Logic Controller (PLC), both in software and hardware~\cite{OpenPLCProject}. It was mainly created for research purposes, because it provides its entire source code~\cite{Alves_openPLC_2014, Alves_OpenPLC_2018}. The current version of the project is OpenPLC V3~\cite{OpenPLCv3}.

    \subsection{Use Case Scenario}
        For this use case, we are going to make the next assumptions:
        \begin{itemize}
            \item The system executing OpenPLC V3 is deployed in an isolated network.
            \item The system running OpenPLC V3 is physically isolated.
            \item The attacker is an insider without access to the systems.
            \item The reference point for the deepness factor will be the \texttt{webserver.py} in Fig.~\ref{fig:edgOpenPLCv3}.
        \end{itemize}

    \subsection{Structure of OpenPLC}
\begin{figure*}[!htb]
          \begin{center}
          \includegraphics[width=0.74\textwidth]{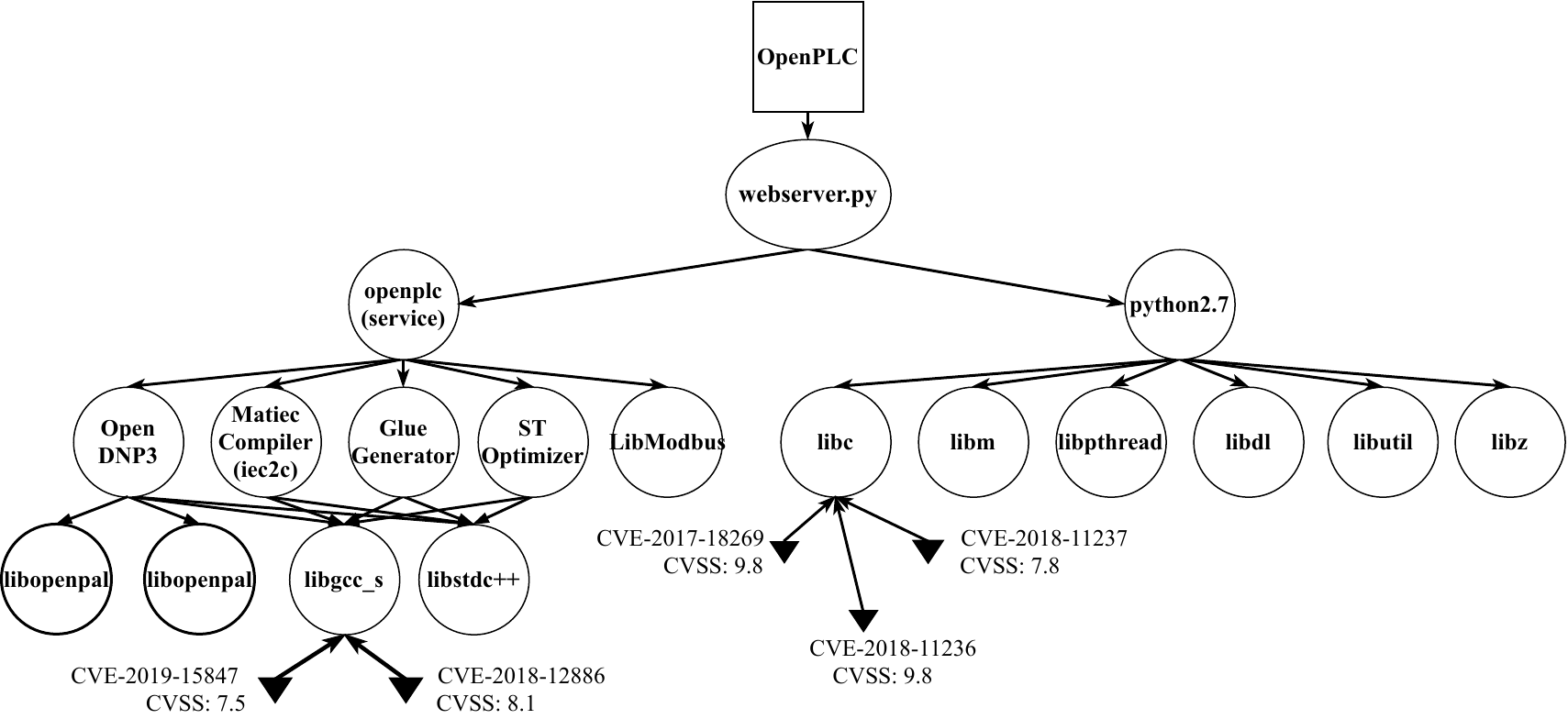}
          \caption{Extended Dependency Graph of OpenPLC V3. Circles represent individual assets, black triangles are the vulnerabilities associated to each asset, and the square represent the entry point to the system, or root node of dependency.}
          \label{fig:edgOpenPLCv3}
          \end{center}
        \end{figure*}

        The first step was to obtain the inner structure of OpenPLC V3 using the Extended Dependency Graph (EDG) proposed in~\cite{Longueira-Romero_EDG_2022}. To simplify the obtained graph, we only represented the shortest path to each node, so the worst case scenario (more accessible from the outside) is considered. The result is shown in Fig.~\ref{fig:edgOpenPLCv3}.

    \subsection{Calculation of the Correcting Factors}
        OpenPLC V3 has five vulnerabilities: two vulnerabilities affecting \texttt{libgcc\_s}, and three vulnerabilities affecting \texttt{libc}. Table~\ref{tab:OpenPLCv3correctionFactors} shows each vulnerability in more detail.

        \begin{table*}[!htb]
            \caption{Vulnerabilities present in OpenPLC V3. For each one, the CVSS is shown, together with their associated Attack Vector (AV), and their correction factors.}
            \label{tab:OpenPLCv3correctionFactors}
            \centering
            \begin{tabularx}{1\textwidth}{Xcccccccc}
                \hline
                CVE             & CVSS & Attack Vector & Functionality ($\rho$) & Deepness ($\beta$) & Context ($\gamma$) & Exploit ($\mu$) & Summarized ($\lambda$) & Corrected CVSS \\ \hline
                CVE-2017-18269  & 9.8  & Network       & 1      & 0.5     & 1        & 1.25  & 0.625    & 6.125 \\
                CVE-2018-11236  & 9.8  & Network       & 0      & 0.5     & 1        & 0     & 0        & 0     \\
                CVE-2018-11237  & 7.8  & Local         & 1      & 0.5     & 0        & 1.25  & 0        & 0     \\
                CVE-2018-12886  & 8.1  & Network       & 1      & 0.25    & 1        & 1.25  & 0.313   & 2.530  \\
                CVE-2019-15847  & 7.5  & Network       & 1      & 0.25    & 1        & 1.25  & 0.313   & 2.344  \\
                \hline
            \end{tabularx}%
        \end{table*}

        From these data, it is possible to obtain all corrections factors for each vulnerability, as follows (Table~\ref{tab:OpenPLCv3correctionFactors} summarizes the results):
 
        \subsubsection{Functionality Factor \texorpdfstring{$(\rho)$}{r}}
            This factor is obtained from the analysis of the description of each CVE. From these data, we have to decide whether the functionality of OpenPLC V3 is affected (``1'') or not (``0'').

        \subsubsection{Deepness Factor \texorpdfstring{$(\beta)$}{b}}
            By taking a look at Fig.~\ref{fig:edgOpenPLCv3}, it can be seen that the maximum deepness level is four. So the possible values for the deepness factor are the ones shown in Fig.~\ref{fig:deepnessFactor}. More precisely, vulnerabilities CVE-2019-15847 and CVE-2018-12886 have a deepness factor of 0.25, because they are at level four. By contrast, vulnerabilities CVE-2017-18269, CVE-2018-11236, and CVE-2018-11237 have a deepness factor of 0.5, because they are at level three.

        \subsubsection{Context Factor \texorpdfstring{$(\gamma)$}{g}}
            From the initial assumptions, insiders can only exploit the existing vulnerabilities from the local network. This means that every vulnerability that has an attack vector of ``network'' (N) can be exploited, thus CVE-2017-18269, CVE-2018-11236, CVE-2018-12886, and CVE-2019-15847 are exploitable by the attacker. Vulnerabilities whose attack vector is ``local'' (L) cannot be exploited, because physical access is needed. Therefore, CVE-2018-11237 cannot be exploited.

        \subsubsection{Exploit Factor \texorpdfstring{$(\mu)$}{m}}
            Public databases have to be queried to find existing exploits for each vulnerability. According to their state of development, a different value is assigned.

        \subsubsection{Summarized Factor \texorpdfstring{$(\lambda)$}{l}}
            The summarized factor for each vulnerability is obtained as the product of the previous factors, as shown in Equation~\ref{eq:summarizedFactor}. At this step, by taking a look at the resulting values of $\lambda$, it is possible to know which CVSS will contribute to the final aggregation and in which percentage ($\lambda > 0$), and which ones will not contribute at all ($\lambda = 0$).

        \subsubsection{Average Factor \texorpdfstring{$(\sigma)$}{s}}
            Finally, we obtained the average factor by calculating the arithmetic mean of all the initial CVSS values: $\sigma = 8.6$.

    \subsection{Aggregation}
        The previous step before the aggregation is obtaining the corrected CVSS value for each initial CVSS. This is done by multiplying each CVSS by their corresponding summarized value ($\lambda$). The corrected values are shown in Table~\ref{tab:OpenPLCv3correctionFactors}.
        
        Finally, the aggregation is performed using the corrected CVSS values. The aggregation is an iterative process that takes the first two values to be added, and adds them using Equation~\ref{eq:addition}. Then, this result is added to the third value to be added, and so on, until there are no more values.
        
        For OpenPLC V3, this process returns a final aggregated value of $9.1$. Without the correction factors, the result would be $10$. Nevertheless, taking into account features such as the exploitability of the vulnerabilities, the context of the SUT, or its functionalities, we can select the most important CVSS values to be aggregated. With such process, the total amount of CVSS values to be added is simplified. This also helps to simplify potential attack paths.
        
        This result was obtained aggregating three of the five CVSS values present in OpenPLC V3. The associated CVSS for CVE-2018-11236 and CVE-2018-11237 were not taking into account for the aggregation, because they do not affect to any functionality of the system, Moreover, CVE-2018-11237 cannot be exploited in the conditions described in the use case.
        
        CVE-2017-18269 (with an associated CVSS of 9.8) is the vulnerability with the highest value for $\lambda$. Therefore, it is going to contribute the most to the final aggregated value. CVE-2018-12886 and CVE-2019-15847 follow with a CVSS of 8.1 and 7.5 respectively. As it is shown, the selected vulnerabilities have a high CVSS, so it is expected that the aggregated value would be also high. This is reflected in the obtained result of $9.1$.
        
        Finally, it is worth highlighting that the final result is lower than the highest CVSS value present in OpenPLC V3. This difference is due to the effect of the correction factors: as the CVE-2017-18269 is further away from the entry point of the system (in layer 3), its real CVSS value in lower.
\section{Conclusions and Future Work}
\label{sec:conclusions}
    In this research work, we proposed a new aggregation algorithm for CVSS values. The proposed approach integrates correction factors to select the most relevant CVSS values to be added based on contextual information. For each vulnerability, we check for:

    \begin{enumerate}
        \item Functionality disruption.
        \item Exploitation difficulty.
        \item Existence of exploits, and their development state.
        \item Context of deployment.
    \end{enumerate}

    We assigned a different correction factor to each one of the previous properties to further ponder the initial CVSS value and adjust it to the real context where the system is operating.

    The proposed aggregation algorithm was applied to OpenPLC V3 in a use case. Two of the existing vulnerabilities were filtered out by the algorithm, as they cannot be exploited in the described context of OpenPLC V3. The rest of the vulnerabilities were aggregated, and the result ($9.1$) was indeed lower than the highest CVSS present in the system ($9.8$). This shows that the CVSS for each vulnerability was correctly adjusted to the real context of deployment of OpenPLC V3.

    As future work, we plan to perform the aggregation at the submetric level of the CVSS, instead of using the base metric value, giving more granular values for each factor.

\section*{Acknowledgements}
    Iñaki Garitano is a member of the Intelligent Systems for Industrial Systems research group at Mondragon Unibertsitatea (IT1676-22), supported by the Department of Education, Universities and Research of the Basque Government. This work was partially supported by the \textit{Ayudas Cervera para Centros Tecnológicos} grant of the Spanish Center for the Development of Industrial Technology (CDTI) under the project EGIDA (CER-20191012), and by the Basque Country Government under the ELKARTEK program, project REMEDY - Real Time Control And Embedded Security (KK-2021/00091).
    \bibliographystyle{ieeetr}
    \bibliography{Bibliography}

\end{document}